\documentstyle[sprocl,psfig]{article}

\def\be{\begin{equation}}
\def\ee{\end{equation}}
\def\bea{\begin{eqnarray}}
\def\eea{\end{eqnarray}}

\begin{document}
\newcommand{\pomeron}{{\protect\rm l\hspace*{-0.15em}P}}

\title{
Fractality of pomeron-exchange processes
in diffractive DIS
}

\author{Zhang Yang}

\address{Institut f\"ur theoretische Physik, FU Berlin\\
 Arnimallee 14, 14195 Berlin, Germany\\E-mail: zhang@physik.fu-berlin.de}


\maketitle\abstracts{ 
By using Monte Carlo simulation of pomeron exchange model,
the dependence of the fractal behavior of the 
pomeron induced system in deep inelastic lepton-nucleon scattering
upon the diffractive kinematic
variables is found rather
robust and not sensitive to the distinct parameterization of the
pomeron flux factor and structure function.
Based on this characteristic fractal plot,
a feasible experimental test of the phenomenological
pomeron-exchange model 
in DESY $ep$ collider HERA is proposed.
}

High energy elastic and diffractive
processes
have long been described by 
using the phenomenology of Regge
theory\cite{regge} by 
the $t$-channel exchange of mesons and,
at high energy, by the leading vacuum singularity, i.e. the
pomeron\cite{pom}.
Because of the ignorance of the nature of 
the pomeron and its reaction mechanisms, there exists different kinds
of approaches and parameterizations of pomeron dynamics
in current Regge theory\cite{para}. And
all the calculation results 
concerning the collective aspects of the diffractive processes,
such as
cross section of hard diffraction\cite{para},
the distribution of large rapidity gap\cite{rapgap},
jet production\cite{para,ua8,hern} in hard diffractive processes etc.
were found very
sensitive to the distinct parameterization of the pomeron.
In this respect, it is natural to ask the following questions:
Is there a way to test and justify the pomeron exchange model 
by using current diffractive experimental equipment
(e.g. DESY $ep$ collider HERA)
while
the criterion to the experimental
measurements does not depend upon concrete parameterization of pomeron?
If yes, 
what is the characteristic behavior of the pomeron exchange
model in the expected experimental measurements?

Having in mind that the fractal and fluctuation
pattern of the multiparticle production
reveals the nature of the correlations 
of the spatial-temporal evolutions
in both levels of parton
and hadron and is, therefore,
sensitive to the interact dynamics of the high-energy
process\cite{bial,inte},
it has been proposed\cite{zhangyang} to
investigate the fractal behavior of the diffractively produced
system by calculating the scaled factorial moments of the
multihadronic final state.
In this talk, we find the dependence of the fractal
behavior of the pomeron induced system upon the
diffractive kinematic
variables is rather robust and not sensitive to 
the different parameterization of the phenomenological pomeron model
in the deep inelastic lepton-nucleon scattering (DIS).
So the characteristic plot about fractal behavior of the
diffractively produced system can be considered 
as a clear experimental test of the
pomeron exchange model within DESY $ep$ collider HERA.

The fractal (or intermittency) behavior of the diffractively produced
system in DIS (and also hadron-hadron collider) can be extracted
by measuring the $q$-order scaled factorial moments (FMs)
of the final-state hadrons excluding the intact proton from the
incident beam, which
are defined by\cite{bial}
\begin{equation}
\label{fm}
F_q(\delta x) = {1\over M}\sum ^M _{m=1}
    {\langle n_m(n_m-1)\dots (n_m-q+1)\rangle \over
    \langle n_m \rangle ^q},
\end{equation}
where, $x$ is some phase space variable of the multihadronic
final-state, e.g. (pseudo-)rapidity, 
the scale of phase space $\delta x=\Delta x/M$
is the bin width for a $M$-partition of the region $\Delta x$ in
consideration, $n_m$ is the multiplicity of diffractively produced 
hadrons in the $m$th bin, $\langle
\cdots\rangle$ denotes the vertical averaging with the different
events for a fixed scale $\delta x$. 

The manifestation of the fractality and intermittency in
high energy multiparticle production refers to the anomalous scaling
behavior of FM\cite{bial,inte}
\begin{equation}
\label{scal}
F_q(\delta x)\sim (\delta x)^{-\phi_q}\sim M^{\phi_q},
       \ \ \ {\rm as}\ M\to \infty,\delta x\to 0.
\end{equation}
The $q$-order intermittency index $\phi_q$ can be connected
with the anomalous fractal dimension $d_q$ of rank $q$ of
spatial-temporal evolution of high energy collisions as\cite{frac}
\begin{equation}
\label{dime}
d_q={\phi _q/(q-1)}
\end{equation}

Pomeron exchange in Regge theory has been used to describe successfully
the main features of the high energy
elastic and diffractive 
process.
While waiting for the experimental measurements of
DESY $ep$ collider HERA (and also hadron-hadron collider)
for the above-mentioned 
fractal behavior of the diffractively producted system,
let us now take a closer look at what
we can learn from the current pomeron exchange theory.

Pomeron factorization allows the diffractive hard scattering
cross-section to be written as
\begin{equation}
\label{kros}
{d^4\sigma (ep\to e+p+X)\over dx_\pomeron
dtd\beta dQ^2}=f_{p\pomeron}(t, x_\pomeron ){d^2\sigma (e\pomeron\to
e+X)\over d\beta dQ^2},
\end{equation}
where $x_B$ and $Q^2$ are the usual deep-inelastic variables;
$x_\pomeron$ is the longitudinal momentum fraction of pomeron emitted
by proton;
and $\beta =x_B/x_\pomeron$.
The first
factor of the right-hand-side of 
Eq. (\ref{kros}) is the pomeron flux, i.e. probability of
emitting a pomeron from the proton;
the second factor, i.e.
lepton-pomeron hard cross section
which is assumed to be independent of 
negative mass-squared
$t$, can be calculated in the
way that 
\begin{equation}
\label{hard}
{d^2\sigma (e\pomeron\to e+X)\over d\beta dQ^2}=\int d\beta' G(\beta' )
{d^2\hat{\sigma}_{\rm hard}(e+{\rm parton}\to e+X)\over d\beta dQ^2},
\end{equation}
if we could figure out
the density $G(\beta' )$ of the quarks and
gluons with fraction $\beta'$ of the pomeron momentum.
Needless to say, hard scattering cross section $d\hat{\sigma}_{\rm
hard}(e+{\rm parton}\to e+X)$ is to be computed 
in perturbation theory
and the Born-level cross section
$d\hat{\sigma}_{\rm hard}$ is proportional to $\delta(\beta-\beta')$.
But both pomeron flux and structure function are still main 
uncertainties in pomeron model,
and it is even unknown
whether the pomeron consist mainly of
gluons\cite{para,ingelmanp} or of quark (see last Donnachie and
Landshoff's papers in Ref. \cite{para}), although
measurements of hard diffractive scattering have been
performed 
in both lepton-hadron and hadron-hadron collider.

In addition to these theoretical uncertainties there is also a
uncertainty in the $Q^2$ evolution of the parton
densities of the pomeron. 
Numerical calculations using ordinary QCD evolution equation 
(Altarelli-Parisi or DGLAP\cite{dglap}), 
and GLR-MQ\cite{mq} equation in which the inverse recombination
processes of partons has been taken into account,
turned 
out that\cite{ingelmanp}
the $Q^2$ evolution of the pomeron structure function can be very much
different depending upon whether the non-linear recombination term 
of the QCD evolution equation is
included or not.
Furthermore,
depending upon the initial parton distribution at a given momentum
scale which is unknown, 
the size of nonlinear term may become too large for the QCD
evolution equation to be reliable without further, but also unknown,
correction.
By assuming both leading and subleading Regge trajectory,
a fit according to the NLO DGLAP evolution equations to HERA
data\cite{h1} of
$F_2^{D(3)}(x_\pomeron,\beta,Q^2)$ has favored a rather peculiar
``one-hard-gluon'' distribution for the pomeron.
Since 
what we try to pursue in this note is to find out whether
and in which
range the
fractal behavior of pomeron induced system depends on varieties of
different parameterization of pomeron, we leave the possible anomalous
scaling behavior in the QCD evolution processes to the further
discussion\cite{zhang1}.

In typical kinematic region of hard diffractive processes of
DESY $ep$ collider HERA (say,
i.e. $M_X>1.1 {\rm GeV}$, and $x_\pomeron <0.1$), 
major properties of the diffractive events
can be well reproduced by RAPGAP generator (see,
e.g. \cite{rapgap,hern,h1}). 
In the following intermittency analysis of the lepton-nucleon
diffractive process, we
use RAPGAP generator\cite{rapgap} to simulate the pomeron exchange
processes, 
in which
the virtual photon $(\gamma^*)$ will interact
directly with a parton constituent of the pomeron for a chosen pomeron 
flux and structure function. In addition to the
$O(\alpha_{em})$ quark-parton model diagram 
$(\gamma^*q\to q)$,
the photon-gluon fusion $(\gamma^*g\to q\bar q)$ and
QCD-Compton $(\gamma^*q\to qg)$ processes are generated
according to the $O(\alpha_{em}\alpha_s)$ matrix elements.
Higher order QCD corrections are provided by the colour dipole
model as implemented in 
(ARIADNE)\cite{ariadne}, and the hadronization is performed using
the JETSET\cite{jetset}. 
The QED radiative processes are included via
an interface to the program HERACLES\cite{heracles}.

\vskip 0.5cm
\begin{figure}[vhb]
\vskip -1.0cm
\centerline{
\psfig{figure=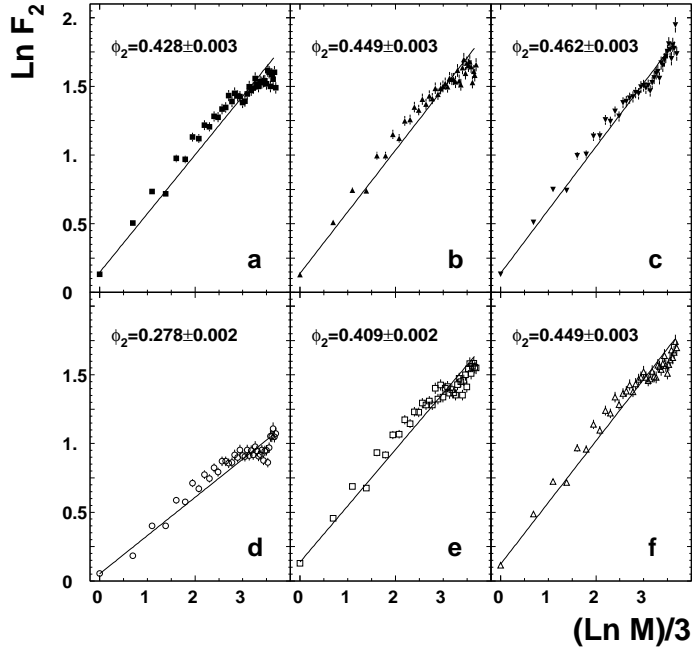,width=10cm}
}
\vskip -1.0cm
\caption{
The second-order scaled factorial moments $F_2$ 
getting from MC simulation of RAPGAP generator{\protect\cite{rapgap}}
versus the number $M$ of
subintervals of 3-dimensional ($\eta,p_\bot,\phi$) phase space in
log-log plot,
and the intermittency index $\phi_2$ correspondingly.
The different kinds of points denote different parameterization of the
pomeron flux factor $f_{p\pomeron}(t.x_\pomeron)$ and structure
function $G(\beta)$, see text for detail.}
\end{figure}
\vskip -0.1cm

By chosing the concrete
pomeron flux factor $f_{p\pomeron}(t,x_\pomeron)$ and
the pomeron structure function $G(\beta)$,
we generate 100,000 MC events,
and calculate the second-order factorial moments
in 3-dimensional ($\eta,p_\bot,\phi$) phase space, where the
pseudorapidity $\eta$, transverse momentum $p_\bot$ and the azimuthal
angle $\phi$ are defined with respect to the sphericity axis of the
event. The cumulative variables $X$ translated from
$x=(\eta,p_\bot,\phi)$, i.e.
\begin{equation}
\label{cumu}
X(x)=\int^x_{x_{\rm min}}\rho (x)dx/\int^{x_{\rm max}}_{x_{\rm min}}
\rho (x)dx,
\end{equation}
were used to rule out the enhancement of FMs from 
a non-uniform inclusive spectrum $\rho (x)$ of the final 
produced particles\cite{bial1}.
The obtained result of second-order FM versus the decreasing scale of
the phase space is shown in Fig.1 in double logarithm.
There exists obviously anomalous scaling behavior in the pomeron
induced interaction, 
so we fit the points in Fig. 1
to Eq. (\ref{scal}) with least square method and
extract the intermittency index $\phi_2$.
In Fig. 1, we have also shown 
the Monte Carlo result of the second-order factorial
moments for different kinds of the parameterization of the pomeron
flux factor and structure function.
In Fig. 1(a), (b), and (c), 
we keep the pomeron structure function $G(\beta)$ fixed
as $\beta G(\beta)=6 \beta (1-\beta)$
but vary the pomeron flux factor $f_{p\pomeron}(t,x_\pomeron)$ 
as 
$f_{p\pomeron}(t, x_\pomeron )={\beta_{p\pomeron}^2(t)\over
16\pi}x_\pomeron^{1-2\alpha_\pomeron (t)}$,
$f_{p\pomeron}(t,x_\pomeron)={1\over 2.3}{1\over x_\pomeron}(6.38
e^{-8|t|}+0.424e^{-3|t|})$ and $f_{p\pomeron}(t,x_\pomeron)={9\beta^2_{q\pomeron}\over
4\pi^2}[F_1(t)]^2x_\pomeron^{1-2\alpha_\pomeron (t)}
$ respectively\cite{para}. The fractal behaviors of the pomeron
induced system keep almost unchanged for different flux factor.
On the contrary we keep the pomeron flux factor
fixed as $f_{p\pomeron}(t, x_\pomeron )={\beta_{p\pomeron}^2(t)\over
16\pi}x_\pomeron^{1-2\alpha_\pomeron (t)}$
in Fig.1(a), (d), (e), and (f), 
but vary the pomeron structure function as 
$\beta G(\beta)=6 \beta (1-\beta)$,
$\beta G(\beta)=6 (1-\beta)^5$,
$\beta G(\beta)=(0.18+5.46\beta )(1-\beta )$ 
and
$\beta G(\beta)=0.077\pi \beta(1-\beta)$
respectively (see Ref.\cite{para} for all these 
parameterisations of pomeron flux and
structure function which are extensively used).
For a given pomeron flux, 
the fractal behaviors become weaker when the pomeron
become softer. In Fig. 1(d) the parton distribution is as soft as that
in proton, the intermittency index is smallest, which is
understandable since
if the hard parton in pomeron is involved
it is more possible to evoke jets 
and then the anomalous short-range correlation
in the final-state 
so that the intermittency index increases, and vice versa.

It is of
the special interest to investigate the dependence of the fractal
behavior of the pomeron induced system upon the  diffractive
kinematic variables.
We generate 500,000 events by RAPGAP Monte
Carlo generator, and divided
the whole sample into 10 subsamples according to the
diffractive kinematic variables, e.g. $x_B$.
For each subsample, 
we calculate the second order scaled FM 
and the intermittency index $\phi_2$,
and to see how the fractal
behavior of the pomeron induced multihadronic final
states depends upon the considered kinematic variable. 
In Fig. 2 is shown the dependence of the second order intermittency
index $\phi_2$ on the different diffractive kinematic variables. 
Since
it is well known that the gluon density increase sharply
as $x_B$ decreases in small-$x_B$
region, the
MC result from pomeron model in Fig. 2(a) means that the anomalous
fractal dimension $d_2$ of the diffractively produced 
system decreases with increasing gluon density,
which is not inconceivable if one takes into account
the fact (see, e.g. \cite{inte,frac})
that the effect of superposition of fractal systems can
remarkably weaken the intermittency of whole system.

\vskip 0.5cm
\begin{figure}[vhb]
\vskip -1.cm
\centerline{
\psfig{figure=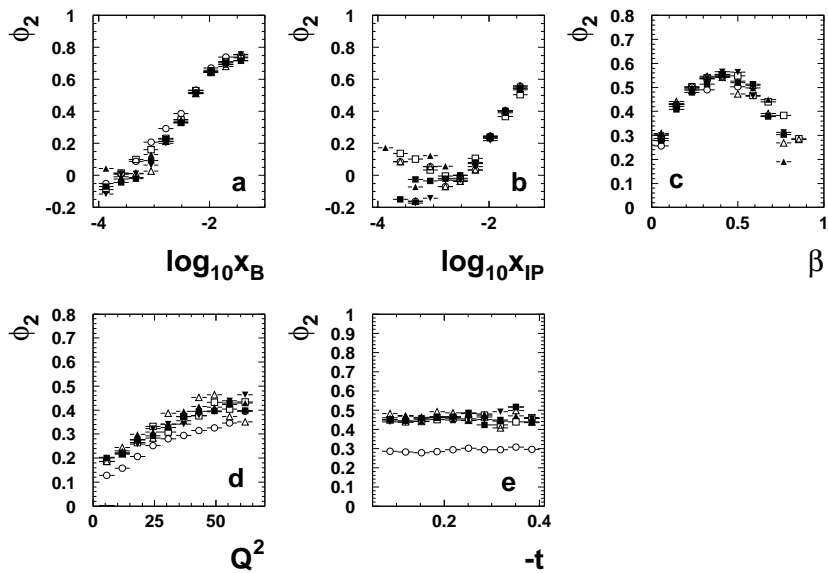,width=12cm}
}
\vskip -2.cm
\caption{
The dependence of second-order intermittency index $\phi_2$ in
RAPGAP{\protect\cite{rapgap}} 
Monte Carlo implementation upon different kinematic variables.
The different shapes of points denote different parameterization of
pomeron flux factors and the structure functions in the same way as that
in Fig. 1.
}
\end{figure}

\vskip -0.1cm

Obvious dependence of $\phi_2$ on 
pomeron momentum fraction
$x_\pomeron$ of a hadron
and parton momentum fraction $\beta$ of a pomeron as shown in
Fig. 2(b) and (c) implies that,
the intermittency calculated here can not be only referred to the
hadronization processes and
there should be substantial
correlations between the
fractal behaviours and pomeron dynamics.
In Fig. 2(a) and (b), the intermittency index $\phi_2$ is less than $0$
for the lower $x_B$ and $x_\pomeron$, which can be imputed to the
constraint of the momentum conservation in the 
high energy
process\cite{liu}. 
Since the Leading Proton
Spectrometer (LPS) has been used in ZEUS detector to detect protons
scattered at very small angles (say, $\leq$ 1 mrad), which make it
possible 
to measure precisely the square of the four-momentum transfer $t$ at
the proton 
vertex, we also showed in Fig. 2(e) the $t$-dependence of
second order intermittency index in the $t$-region of LPS
detector, i.e. $0.07<-t<0.4\  {\rm GeV}^2$.
To be different from the results of other kinematic variables, the
fractal index for the $\gamma^*\pomeron$ system doesn't depend
upon the $t$.

Especially, we calculate the intermittency index 
shown in Fig. 2
using different kinds of pomeron parameterization. 
We denote the different shapes of points in Fig. 2
for the different kinds of the pomeron parameterization,
just in the same way as that in Fig. 1.
Just as mentioned above,
in conventional investigation, 
the pomeron theory has been used to 
compare with the data about cross
section of hard diffraction\cite{para}, the rapidity
distribution of large rapidity gap\cite{rapgap},
and jet rapidity distribution\cite{para,hern} and
jet shape\cite{ua8,hern} etc., 
where the results of the pomeron model 
concerning these collective nature of diffractive
process
were found very
sensitive to the parameterization of the pomeron, and the experimental
data in different aspects preferred different kinds of
parameterization\cite{ua8,hern,zeussh,h1}.
It is remarkable
that the dependence of
the intermittency index, 
which concerned with the {\it inherent} scaling behaviors
of diffractive processes,
upon the diffractive kinematic variables
in this implementation of the pomeron exchange model are
rather robust and
almost the same for the different parameterization of the pomeron flux
and structure function!

Having in mind that the multiplicity measurement of multihadronic
production is available in DESY $ep$ collider HERA and especially
the multiplicity moments and KNO scaling behaviours
of final hadrons in deep-inelastic processes have been
already studied in recent years\cite{multi},
it is feasible and urgent to check this characteristic
fractal plot in DESY $ep$ collider HERA.
Substantial revision would be necessary
in the manner in which we
have treated diffraction 
if it should turn out that experimental
measurements differ drastically from this characteristic plot
presented here.

\section*{Acknowledgments}
I would like to thank T. Meng, R. Rittel and K. Tabelow for the
helpful discussion,
H. Jung for correspondence,
and the Alexander von Humboldt Stiftung for
financial support.
Thanks are also due to W. Kittel for the invitation and arrangement of
my talk, and his organizing such an enjoyable session.
Because of the efforts of N.G. Antoniou, L. Kontraros
and the other members of the organizing committee, the symposium was
extremely well organized.

\end{document}